\documentclass[conference]{IEEEtran}
\IEEEoverridecommandlockouts

\usepackage{todonotes}
\usepackage{cite}
\usepackage{amsmath,amssymb,amsfonts}
\usepackage{algorithmic}
\usepackage{graphicx}
\usepackage{textcomp}
\usepackage{xcolor}
\usepackage[acronym]{glossaries}

\def\BibTeX{{\rm B\kern-.05em{\sc i\kern-.025em b}\kern-.08em
    T\kern-.1667em\lower.7ex\hbox{E}\kern-.125emX}}

\newacronym{dm}{DM}{Decision Making}
\newacronym{mcc}{MCC}{Mobile Cloud Computing}
\newacronym{mec}{MEC}{Mobile Edge Computing}
\newacronym{cec}{CEC}{Cloud Edge Computing}
\newacronym{rsu}{RSU}{Road Side Unit}
\newacronym{uds}{UDS}{Unified Diagnostic Services}
\newacronym{sovd}{SOVD}{Service-Oriented Vehicle Diagnostics}
\newacronym{ecu}{ecu}{Electronic Control Unit}
\newacronym{dtc}{DTC}{diagnostic trouble codes}
\newacronym{obd}{OBD}{On-Board Diagnostics}
\newacronym{sae}{SAE}{Society of Automotive Engineers}
\newacronym{can}{CAN}{Controller-Area Network}
\newacronym{iso}{ISO}{International Organization for Standardization}
\newacronym{api}{API}{Application Programming Interface}
\newacronym{ugv}{UGV}{Unmanned Ground Vehicle}
\newacronym{sdv}{SDV}{Software Defined Vehicle}

\usepackage[left=0.68in,right=0.673in,top=0.75in,bottom=1.05in]{geometry}

\begin{document}

\title{SDVDiag: A Modular Platform for the Diagnosis of Connected Vehicle Functions\\
}

\author{
\IEEEauthorblockN{Matthias Weiß, Falk Dettinger, and Michael Weyrich}
\IEEEauthorblockA{
\textit{Institute of Industrial Automation and Software (IAS)} \\
\textit{University of Stuttgart} \\
Pfaffenwaldring 47, 70550 Stuttgart, Germany \\
E-Mail: \{matthias.weiss, falk.dettinger, michael.weyrich\}@ias.uni-stuttgart.de}}

\maketitle

\begin{abstract}
Connected and software-defined vehicles promise to offer a broad range of services and advanced functions to customers, aiming to increase passenger comfort and support autonomous driving capabilities. Due to the high reliability and availability requirements of connected vehicles, it is crucial to resolve any occurring failures quickly. To achieve this however, a complex cloud/edge architecture with a mesh of dependencies must be navigated to diagnose the responsible root cause. As such, manual analyses become unfeasible since they would significantly delay the troubleshooting.

To address this challenge, this paper presents SDVDiag, an extensible platform for the automated diagnosis of connected vehicle functions. The platform enables the creation of pipelines that cover all steps from initial data collection to the tracing of potential root causes. In addition, SDVDiag supports self-adaptive behavior by the ability to exchange modules at runtime. Dependencies between functions are detected and continuously updated, resulting in a dynamic graph view of the system. In addition, vital system metrics are monitored for anomalies. Whenever an incident is investigated, a snapshot of the graph is taken and augmented by relevant anomalies. Finally, the analysis is performed by traversing the graph and creating a ranking of the most likely causes.

To evaluate the platform, it is deployed inside an 5G test fleet environment for connected vehicle functions. The results show that injected faults can be detected reliably. As such, the platform offers the potential to gain new insights and reduce downtime by identifying problems and their causes at an early stage.

\end{abstract}


\section{Introduction}
The automotive industry is undergoing a fundamental transformation, shifting from traditionally mechanical systems toward connected and software-defined vehicles (SDVs) \cite{baumann2024connected} \cite{weiss2023continuous}. These vehicles increasingly rely on complex software stacks, continuous over-the-air updates, and cloud-based services to deliver advanced functions such as predictive maintenance, real-time traffic routing, and autonomous driving support. As these systems become more sophisticated, so do the expectations placed upon them—drivers and passengers alike demand seamless user experiences, minimal downtime, and robust reliability even under dynamic operating conditions.

To ensure these expectations are met, connected vehicle platforms must be closely monitored and continuously evaluated. This is typically achieved by establishing data loops that capture operational data across vehicle components, edge nodes, and backend cloud systems \cite{weiss2024ad}. These loops form the basis of observability pipelines that provide valuable insights into system health, service availability, and anomalous behavior. However, the increasing complexity and distribution of these systems introduce significant challenges.
                     
Diagnosing issues in such an environment is far from trivial. When a function degrades or fails, the root cause may lie deep within a chain of dependent services spread across in-vehicle software, edge processing units, and cloud-native components. Manual diagnosis in this context is not only time-consuming but also prone to human error, often requiring engineers to navigate layers of telemetry, logs, and metrics without a unified view of the system’s internal state. This latency in troubleshooting can lead to prolonged downtimes and diminished user trust—both critical concerns for the success of SDVs \cite{weiss2022devops}.

Given these challenges, a pressing research question arises:
\textit{How can an automated and scalable diagnosis of connected vehicle functions across complex, distributed environments be enabled?}
This paper addresses this question by introducing SDVDiag, a flexible and extensible platform designed to automate the end-to-end process of diagnosing faults in connected vehicle systems.

The remainder of this paper is structured as follows: Section II and Secton III provide an overview of related work with regards to distributed automotive systems and observability. Section IV introduces the architecture of the SDVDiag platform and outlines the automated diagnosis, including graph creation, incident analysis and self-adaptation. Section V presents the evaluation within a 5G test fleet environment and discusses the effectiveness of SDVDiag in identifying injected faults. Finally, Section VI concludes the paper and outlines directions for future research.

\section{Background}
\subsection{Distributed Systems in Vehicular Environments}
The increasing automation and benefits for autonomous navigation, such as cooperative map generation or driving, highlight the need for powerful and scalable backend systems. Typically, both cloud servers and edge clusters are used. While the cloud offers scalability and high computing power \cite{dettinger2024function}, its greater distance to vehicles results in higher latency, making it unsuitable for time-critical applications. In contrast, edge servers, mounted on \gls{rsu}s or cell towers, provide local access to computing resources for latency-critical applications but have limited processing capacity compared to the cloud.

Specific implementations are enabled through the alternative or combined use of cloud and edge. Here, the terms \gls{mcc} \cite{akherfi2018mobile}, \gls{mec} \cite{mao2017survey}, and, in hybrid systems, \gls{cec} \cite{wang2020survey} are significant. \gls{mcc} refers to systems that solely rely on the cloud data center for computation, while \gls{mec} corresponds to the use of edge servers. \gls{cec}, as a hybrid approach, attempts to combine the advantages and disadvantages of \gls{mcc} and \gls{mec}, enabling both the scalability of the cloud and the processing of latency-critical applications on the edge.

Applications in this context are implemented as services that include all dependencies and can be deployed in a scalable manner on cloud and edge servers \cite{peltonen2020edge}, then retrieved by the vehicle. During deployment, reliability and availability of the services are ensured, which is why they are often implemented as distributed applications within computing clusters, such as Kubernetes.

\subsection{Vehicle Data Loops and Observability}
In connected vehicle environments, vast amounts of data are continuously collected to facilitate collaborative use cases, enhance existing software models, and monitor system health to ensure reliability and availability. This systematic collection, processing, and utilization of vehicle-generated data are commonly described as a data loop. Typically, a data loop comprises multiple interconnected stages: data acquisition from vehicles, transmission to backend infrastructure, processing and analysis, model updates or decision-making based on analysis results, and finally, deployment of the updated models or corrective actions back into the vehicle fleet.

Given the distributed nature of connected vehicle systems, which often span both cloud and edge computing environments, specialized methods and tools are essential for effectively managing and analyzing the collected data. A crucial concept in this context is observability, defined as the ability to understand and diagnose the internal state of a system solely based on its external outputs, such as metrics, logs, and traces \cite{sridharan2018distributed, niedermaier2019observability}. Observability enables system engineers to reconstruct the state and behavior of complex systems, thus making it possible to pinpoint the root causes of observed failures quickly and accurately.

To achieve observability in distributed vehicular environments, various tools and practices have been established. These tools typically capture three primary types of data: metrics that quantify the system's state numerically (e.g., CPU load, memory usage, response times), logs that record discrete events and state changes, and traces that document the path of requests across different services \cite{sigelman2010dapper, li2022enjoy}. Together, these data types provide comprehensive visibility into the functioning of vehicle systems and their backend infrastructures.

For the analysis of such observability data, anomaly detection methods are widely employed. Anomaly detection identifies deviations from expected patterns, highlighting potential problems and enabling proactive maintenance. Techniques can broadly be categorized into statistical methods, machine learning-based approaches, and hybrid models. Statistical methods, such as Z-score analysis or seasonal decomposition of time series (e.g., STL), rely on predefined assumptions about data distributions \cite{chandola2009anomaly}. Machine learning-based methods, including supervised approaches like isolation forests and unsupervised models like autoencoders, are particularly prevalent in productive environments due to their effectiveness in detecting complex anomalies without explicit definitions \cite{ruff2021unifying}. However, anomaly detection models employed in production are typically trained statically, necessitating periodic retraining to maintain accuracy as data characteristics evolve.

\section{State of the Art}
\subsection{Vehicle Diagnostics}

For traditional, non-connected cars, \gls{uds} is a widely adopted standard in vehicle diagnostics, serving as a key communication protocol between diagnostic tools and a vehicle's \gls{ecu}s \cite{iso2020uds}. It provides standardized processes for fault detection, analysis, and resolution, helping to identify operational anomalies. In addition to \gls{uds}, several well-established diagnostic protocols contribute to anomaly detection in vehicles, including \gls{obd}-II  \cite{geotab2023obd} and \gls{sae} J1939 \cite{sae2023j13939}.

As vehicle architectures increasingly integrate high performance computing (HPC) units and methods for hardware abstraction, service-oriented approaches are gaining prominence alongside traditional diagnostic protocols. \gls{sovd} is a key innovation designed to address the increasing complexity of software-defined and connected vehicle architectures \cite{asam2022sovd}. By leveraging standardized APIs, \gls{sovd} enables real-time data retrieval, flexible system monitoring, and seamless integration into cloud-based diagnostic frameworks. Yet, to the best of our knowledge, to this date there exists no such diagnostic framework for distributed, connected vehicle functions, which incorporates both vehicle and cloud services.

\subsection{Causal Inference in Distributed Systems} \label{causal}
The increasing complexity and scale of modern distributed systems cause conventional analysis methods to become impractical. Thus automated methods have gained significant attention over the past years, typically involving a two-step process: anomaly detection to identify unexpected system behavior and subsequent causal inference to pinpoint underlying root causes \cite{li2021observability}. For the latter, graph-based methods have become particularly prevalent due to their ability to intuitively represent relationships within distributed systems. Two principal types of graphs are commonly utilized: dependency graphs and causal models \cite{wang2024rca}. Dependency graphs explicitly represent relationships among system components, generally based on observed interactions such as communication patterns or resource usage. Causal models, on the other hand, explicitly capture causal relationships between components, quantifying how changes or anomalies in one component influence others.

To generate causal models, causal discovery methods have been developed, aiming to infer causal relationships directly from observational data. One example that will also be used in this paper is the Amortized Causal Discovery (ACD) framework proposed by Loewe et al. \cite{lowe2022amortized}, which leverages deep learning to efficiently infer causal structures from time series data, particularly suited for dynamic and noisy distributed environments. Despite substantial advancements, the application of current causal inference approaches to connected vehicle environments requires careful scrutiny and extensive adjustments. Notably, these systems present high dynamicity due to frequent software updates and environmental changes, significant heterogeneity across various vehicle functions, and complex integrations spanning vehicle-edge-cloud infrastructures. Addressing these gaps represents a pivotal research direction to enhance reliability and maintainability within connected vehicle environments.

\section{Concept of SDVDiag}
\subsection{Overview}
To deal with the intricate challenges of connected vehicle environments, this paper introduces the concepts of distributed tracing and causality mining to the automotive domain and proposes additional approaches to improve reliability in complex distributed systems. In particular, the following requirements to the concept have been isolated:
\begin{enumerate}
    \item \textbf{Real-time Data Collection}: The platform must continuously collect data from vehicles and backend services in real time to reflect the current system state.
    \item \textbf{Distributed Tracing Capability}: SDVDiag should provide distributed tracing to effectively track requests through various services and infrastructure layers.    
    \item \textbf{Dynamic Dependency Mapping}: The platform must dynamically detect and maintain an updated map of dependencies between different vehicle functions and services.
    \item \textbf{Anomaly Detection}: Robust anomaly detection techniques must identify deviations in metrics to proactively recognize system issues.
    \item \textbf{Causality Analysis and Root-cause Identification}: SDVDiag must perform causality mining to trace relationships between anomalies, identifying potential root causes.
    \item \textbf{Modularity, Extensibility, Scalability}: The platform architecture should support modularity, allowing runtime component exchange, extension, and scaling.
    \item \textbf{Self-adaptation and Continuous Model Updating}: The platform must continuously update causality mining and anomaly detection models to ensure consistent performance even under concept drift.
    \item \textbf{Intuitive Visualization and Reporting}: Diagnostic outcomes, including anomalies and causality graphs, should be presented through clear and intuitive visualizations.
\end{enumerate}

\begin{figure*}[ht]
    \centering
    \includegraphics[width=\linewidth]{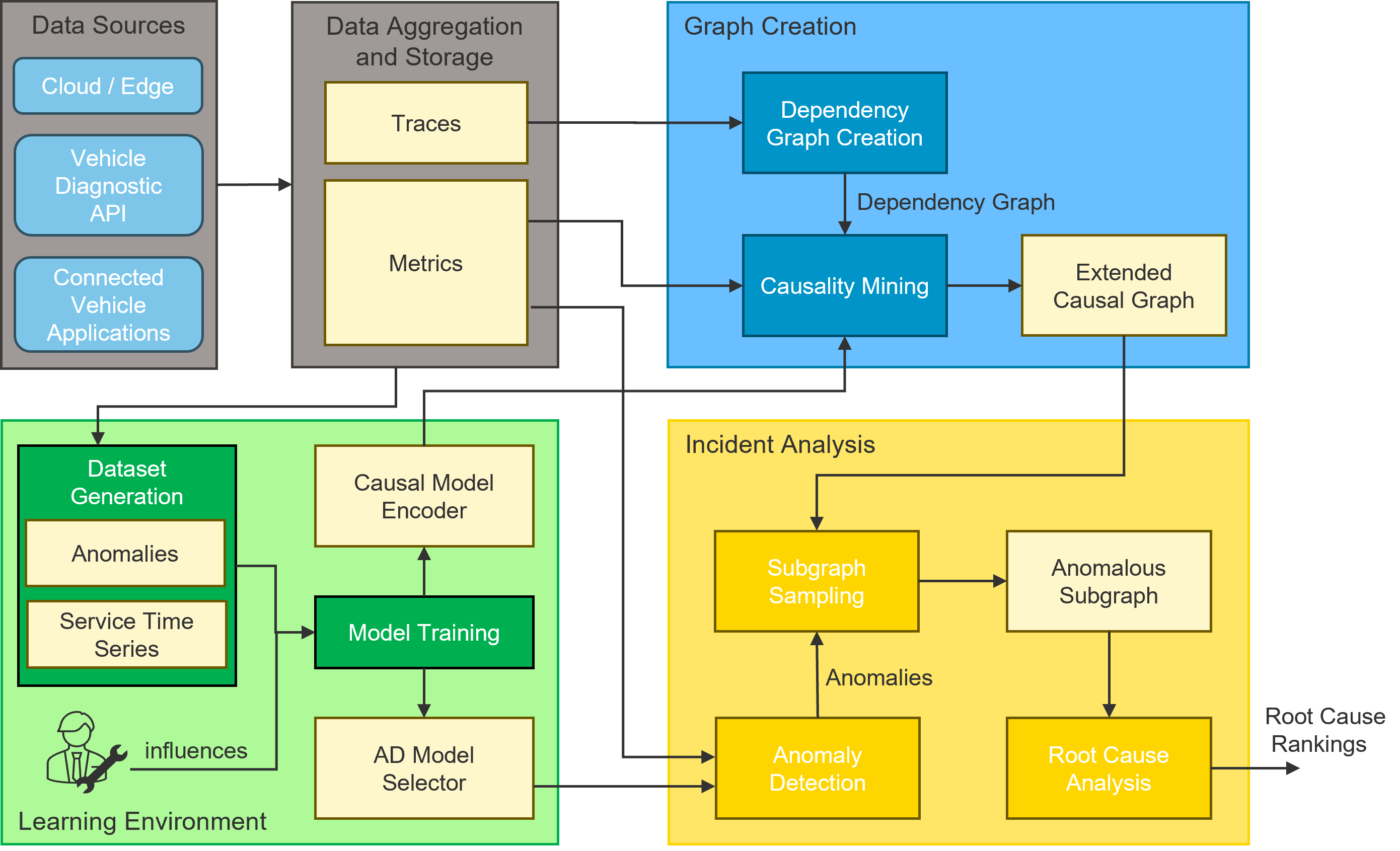}
    \caption{Conceptual Overview of SDVDiag. The platform can roughly be divided in subsystems for the generation of graphs, the analysis of these graphs for detected incidents and a learning environment, in which the used models evolve continuously.}
    \label{fig:concept-overview}
\end{figure*}

To address these requirements, SDVDiag, which is facilitated by a service-based architecture, is designed. Fig. \ref{fig:concept-overview} presents the resulting application and its key subsystems. In general, the architecture can roughly be divided into:

\begin{itemize}
    \item A \textbf{data aggregation and storage} layer (grey), which includes methods and technologies for accessing the information required for the analysis from all involved data sources,
    \item the process of \textbf{graph creation} (blue), which is responsible for generating a comprehensive system overview based on the given data, which can subsequently be used when investigating incidents,
    \item a \textbf{learning environment} (green), in which the involved models are trained continuously based on most recent data to ensure a robust analysis process, and
    \item the actual \textbf{incident analysis} (yellow), where anomalies become causally linked and the most probable root causes are determined.
\end{itemize}

More details on these components are provided in the following.

\subsection{Data Aggregation and Storage}
Foundational Layer that holds all data that is relevant for the continual diagnosis. Data must be collected from all sources (i.e., vehicles, edge nodes and the cloud), stored and preprocessed within acceptable time constraints. To accomplish this, the platform provides support for most state of the art monitoring and observability protocols as they are used in productive environments. In the implemented examples, OpenTelemetry, Apache Kafka and the Delta Lake framework have all been used for data collection and storage in real-time. For the correct operation of the analysis platform, two specific data types must be collected: Traces, by which the data and communication flow between components can be measured, and metrics, which provide information about the system status and performance, such as the resource consumption on available compute nodes. The integration of additional data types like logs or application-specific data is supported as well via the provided platform interfaces.

\subsection{Graph Creation} \label{graph}
This layer is responsible for generating and maintaining a comprehensive overview of the system using a graph-based data model. In this model, nodes represent system components along with their characteristics, while edges symbolize the relationships between these components. SDVDiag uses two distinct types of graphs (also ref. to Section \ref{causal}), whose elements are explained in the following:

\subsubsection{Dependency Graphs}
In distributed systems, runtime information forms the foundation of all analysis processes due to the dynamic behavior of the environment. For example, dependencies between services and instances must be derived from live communication flows, as they are not determined statically at time of development. As such, SDVDiag integrates tools for distributed tracing, by which dependency graphs are constructed, and saves the result in a graph database for further analysis. Additionally, SDVDiag supports to extend the dependency graph to enable a dedicated analysis of specific applications or types of incidents. For the purpose of this paper, timestamps of the most recent recorded communication between two services are added as an example in order to provide context on failure propagation to the analysis. Other possible information to integrate includes current performance metrics and additional knowledge provided by engineers, all of which require compatible causality mining techniques.

\subsubsection{Causal Graphs}
\begin{figure}[htb]
    \centering
    \includegraphics[width=0.95\linewidth]{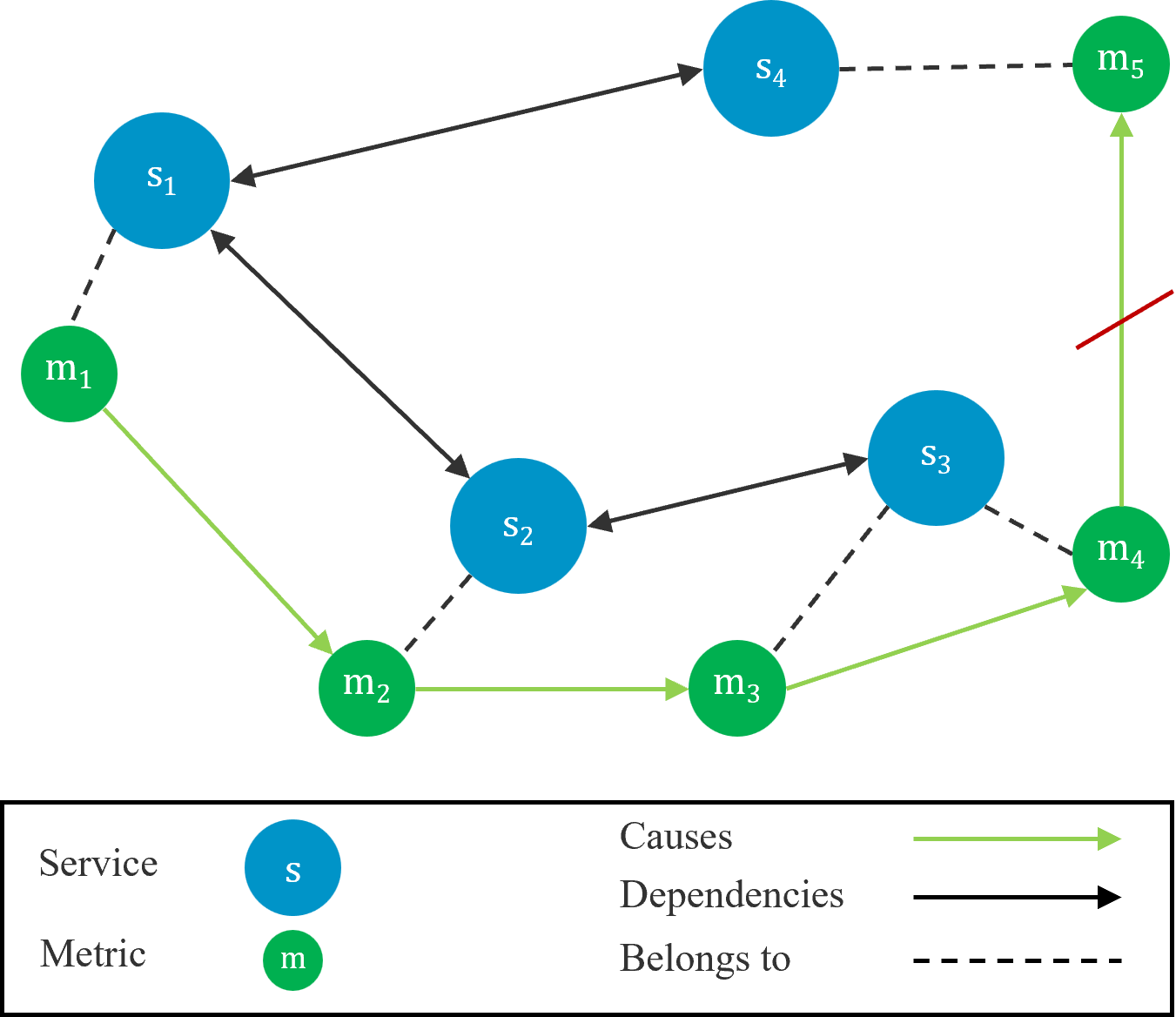}
    \caption{Extended causality graph of SDVDiag. Information about service dependencies is combined with causal relationships, resulting in a comprehensive system overview. Causalities are pruned when there is no direct dependency between two nodes.}
    \label{fig:graph}
\end{figure}

As mentioned in Section \ref{causal}, causal relationships are essential for automated analyses since they quantify the strength of causality between components. To derive these relationships, SDVDiag supports integrating various causal discovery models, which can be exchanged dynamically at runtime due to the modular architecture of the platform. In the context of this paper, the Amortized Causal Discovery (ACD) framework (see Section \ref{causal}) has been incorporated into the platform to identify causal relationships between service instances. During operation, ACD identifies causal relations among monitored performance metrics and constructs an initial causal graph structure, which currently includes metric nodes interconnected by weighted edges representing causal strength. Subsequently, SDVDiag combines this causal information with the existing dependency graph, resulting in an extended causal graph, as illustrated in Figure \ref{fig:graph}. In this combined graph, metric nodes are assigned to their corresponding service instances. Furthermore, to enhance analytical stability, causal edges between metrics are pruned if the associated services or instances are not related according to the dependency graph. The resulting graph serves as the basis for subsequent incident analyses and is continuously updated to reflect the latest system state.

\subsection{Learning Environment} \label{le}
To ensure the adaptability of the analysis platform, the involved models must be continuously updated to account for concept drift (e.g., introduced by software updates) or dynamic changes in system resources. To facilitate this, SDVDiag includes a dedicated learning environment that supports system engineers in generating suitable datasets and controlling the training process of the models. Specifically, two primary models require ongoing updates:

\subsubsection{Causal Model Encoder} 
As mentioned in Section \ref{graph}, the causality mining process requires a model that learns the causal behavior of the system. During operation, this model computes the causal weights between system components, determining the strength of causal relationships between pairs of components. For the purpose of this paper, a model for Amortized Causal Discovery has been trained within the learning environment. The model learns system dynamics based on time series data extracted from operational data collections. A significant limitation of many causal models, including ACD, is that even minor system changes can rapidly degrade the quality of results. To address this issue, retraining of the model is initiated whenever such a change occurs—typically after software updates or deployment modifications. Nevertheless, given the frequent changes typical in complex distributed systems, this limitation remains challenging for conventional causal discovery methods. Consequently, SDVDiag integrates an additional feedback loop, enabling more effective interventions by system engineers and incremental model adjustments that facilitate faster learning during live operation. The detailed concept behind this approach is beyond the scope of this paper and will be elaborated upon in future work.

\subsubsection{Anomaly Detection Model Selector}
To ensure robust anomaly detection, SDVDiag implements automated anomaly detector selection based on the characteristics of individual time series. During operation, features are extracted from each metric, upon which a suitable anomaly detection model is selected, trained, and deployed. This selection is facilitated through a continuous training loop, wherein an agent learns to select the optimal model from a predefined pool based on the extracted features of each time series. For the scope of this paper, both supervised and unsupervised anomaly detection models have been integrated into the available solution space. Initially, the agent undergoes training using historical, labeled datasets to ensure optimal performance. The learning environment supports system engineers in creating these labeled datasets by automatically recording historical data and suggesting labels. Once the initial training phase is complete, the agent can be continuously fine-tuned, which is performed on live data during operation and relies on a feedback mechanism wherein system engineers indicate whether the selected anomaly detection model's assessment for a given time series was correct. The detailed design and evaluation of this loop will be covered in future work.

\subsection{Incident Analysis}
Given the availability of adaptive models (Section \ref{le}) and the extended causal graph (Section \ref{graph}), SDVDiag can proceed to analyze specific system incidents. Incident analysis can either be initiated manually by system engineers or triggered automatically through anomaly detection.

Regardless of the triggering method, SDVDiag subsequently creates a snapshot of the most recent system state for detailed investigation. This snapshot includes the current extended causal graph along with all anomalies detected within a configurable timeframe. Interfaces are provided for additional modifications to this snapshot. For instance, this paper demonstrates a module that prunes graph paths involving services without any detected anomalies, as such services are unlikely to contribute to the investigated incident. Further ML-based sampling techniques have also been developed and can be selected within the platform.

After preparing the subgraph, the actual root-cause analysis is performed. SDVDiag provides modules supporting both first-order and second-order random walks, as well as conventional Fault Tree Analysis (FTA). In this paper, the first-order random walk is used as the default method due to its efficiency in smaller distributed environments. However, the authors suggest to rely on the other options in larger systems with long causal chains. For the first-order walk, system engineers can choose whether the analysis should identify the root cause of a single anomaly or all detected anomalies collectively. When analyzing a single anomaly, the random walk algorithm begins at the node representing the anomaly's metric and traverses the graph along causal relationships. Neighboring nodes are selected randomly, weighted by the strength of the causal relationships—meaning nodes with stronger causal connections are visited more frequently. For analyzing multiple anomalies, the process repeats with different start nodes corresponding to each anomaly, until all anomalies have been processed. Finally, the visit counts for each node are summed and ranked in descending order, providing the system engineer with an ordered list of the most probable root causes for further investigation.

\section{Evaluation}

\begin{figure}[htb]
    \centering
    \includegraphics[width=\linewidth]{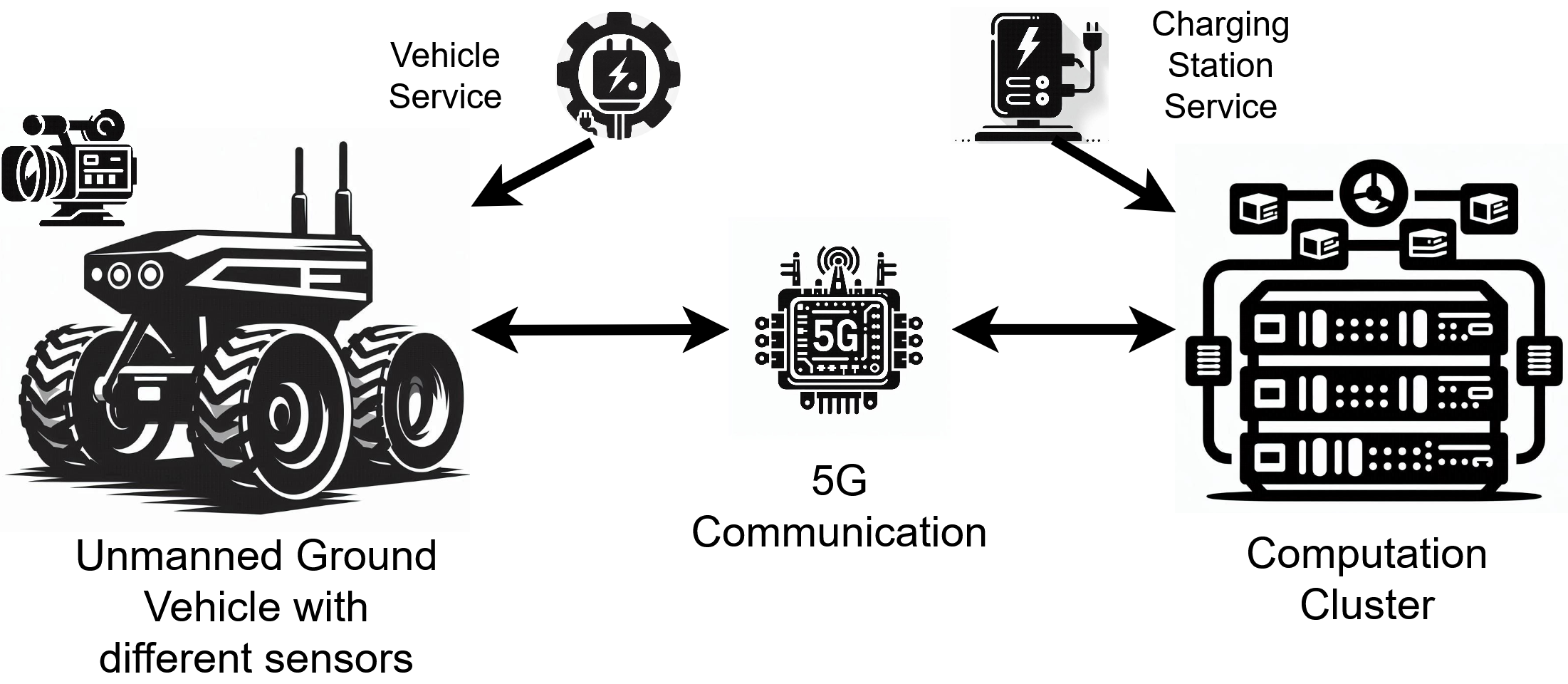}
    \caption{System overview consisting of Unmanned Ground Vehicle with defined sensor set, connected via 5G with a computation cluster providing a charger station service}
    \label{fig:evaluation_overview}
\end{figure}

To evaluate the analysis capabilities of the SDVDiag platform, a smart charging application was developed and deployed within the 5G vehicle test track at the University of Stuttgart. Fig.~\ref{fig:evaluation_overview} provides an overview of the entire system. The hardware setup includes multiple Unmanned Ground Vehicles (UGVs) equipped with various sensors for environmental perception, such as LiDAR and depth cameras, enabling autonomous navigation. Designated parking spots with integrated charging stations are available throughout the track. For efficient data transmission, each UGV is connected to a local, scalable computing cluster via 5G. The cluster, implemented using Kubernetes, consists of one control node and three worker nodes.

Multiple instances of the charging station service are deployed on the cluster, providing real-time access to charging infrastructure data~\cite{weiss2024simulating}. These services process publicly available information to offer key details such as the location, operator, and capacity of charging stations. A REST API enables vehicles to retrieve nearby stations, check availability, and perform spatial queries using geometric distance calculations. The vehicle service allows seamless access to this information by responding to vehicle requests for nearby charging options.

\begin{figure}[htb]
    \centering
    \includegraphics[width=\linewidth]{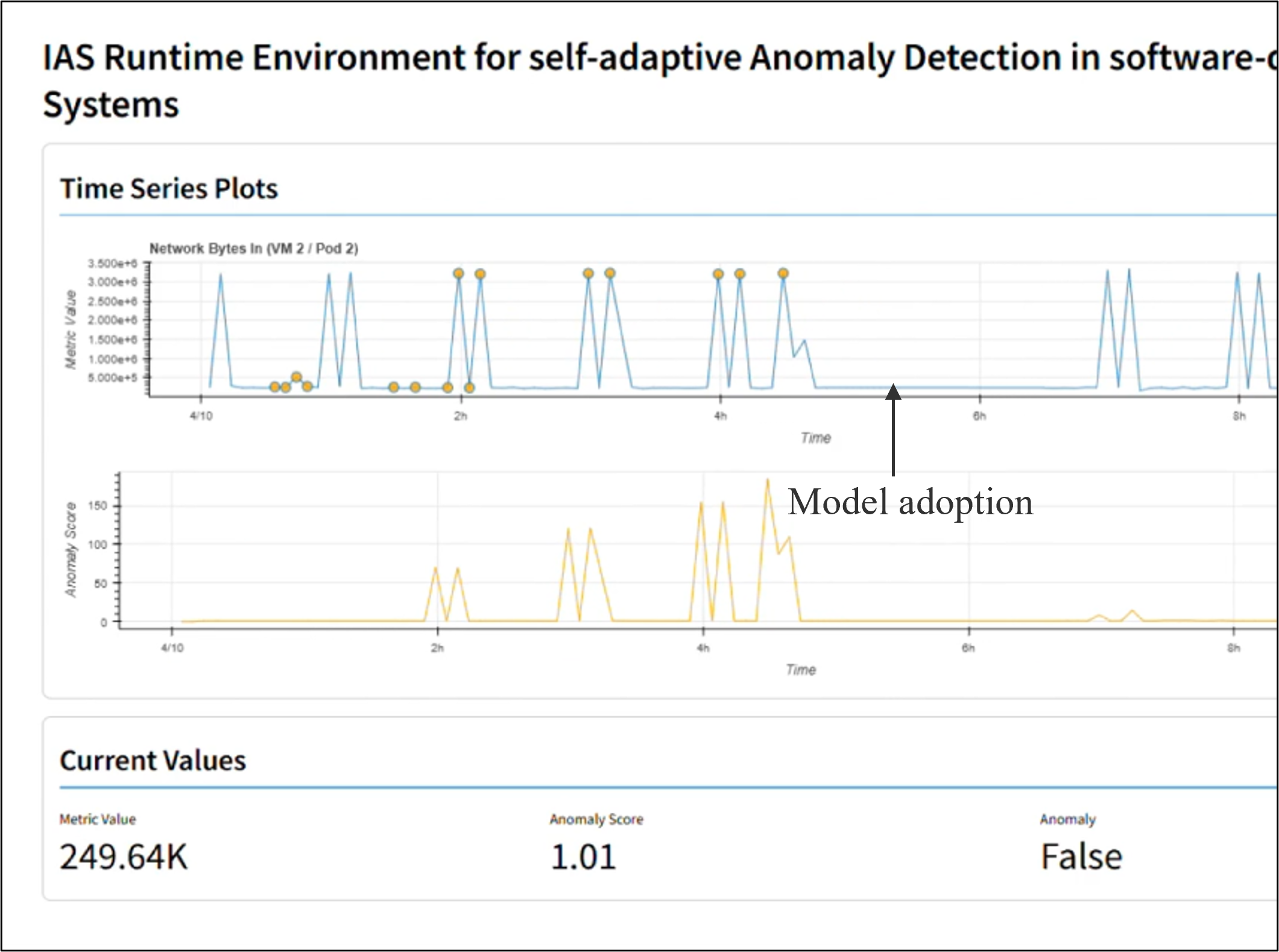}
    \caption{UI of SDVDiag for Anomaly Detection. Yellow dots symbolize detected anomalies. Detection is improved over time by the learning loop.}
    \label{fig:ui}
\end{figure}

To enable analysis in this environment, SDVDiag was integrated to monitor both the Kubernetes cluster and the deployed vehicle services. Traces are collected to construct a dependency graph, while causal discovery is applied to the collected metrics to infer causal relationships. The Causal Model Encoder and the AD Model Selector were trained on historical data and subsequently deployed into the live system. Fig.~\ref{fig:ui} displays the SDVDiag user interface during anomaly detection. The highlighted metric is the transmitted data volume of a specific worker node, which shows characteristic spikes whenever vehicle requests are processed. Anomalies (marked as yellow dots) appear at the beginning of the recording but are identifiable as false positives. As more data becomes available, the AD Model Selector improves its model selection and training, resulting in a noticeable reduction of false positives over time.

Regarding the evaluation of the incident analysis, a scenario was simulated in which a single charging station service experiences increased CPU usage. This anomaly produces several observable effects in the system, including:
\begin{itemize}
    \item A decrease in CPU usage for other charging station services on the same worker node due to limited available resources.
    \item An increase in CPU usage for charging station services on other nodes, as requests are rerouted to available services.
    \item Increased CPU usage for vehicle services interacting with the affected charging station, as they eventually assume the service is unreachable and attempt to compute the closest alternative locally.
\end{itemize}

\begin{figure}[b]
    \centering
    \includegraphics[width=\linewidth]{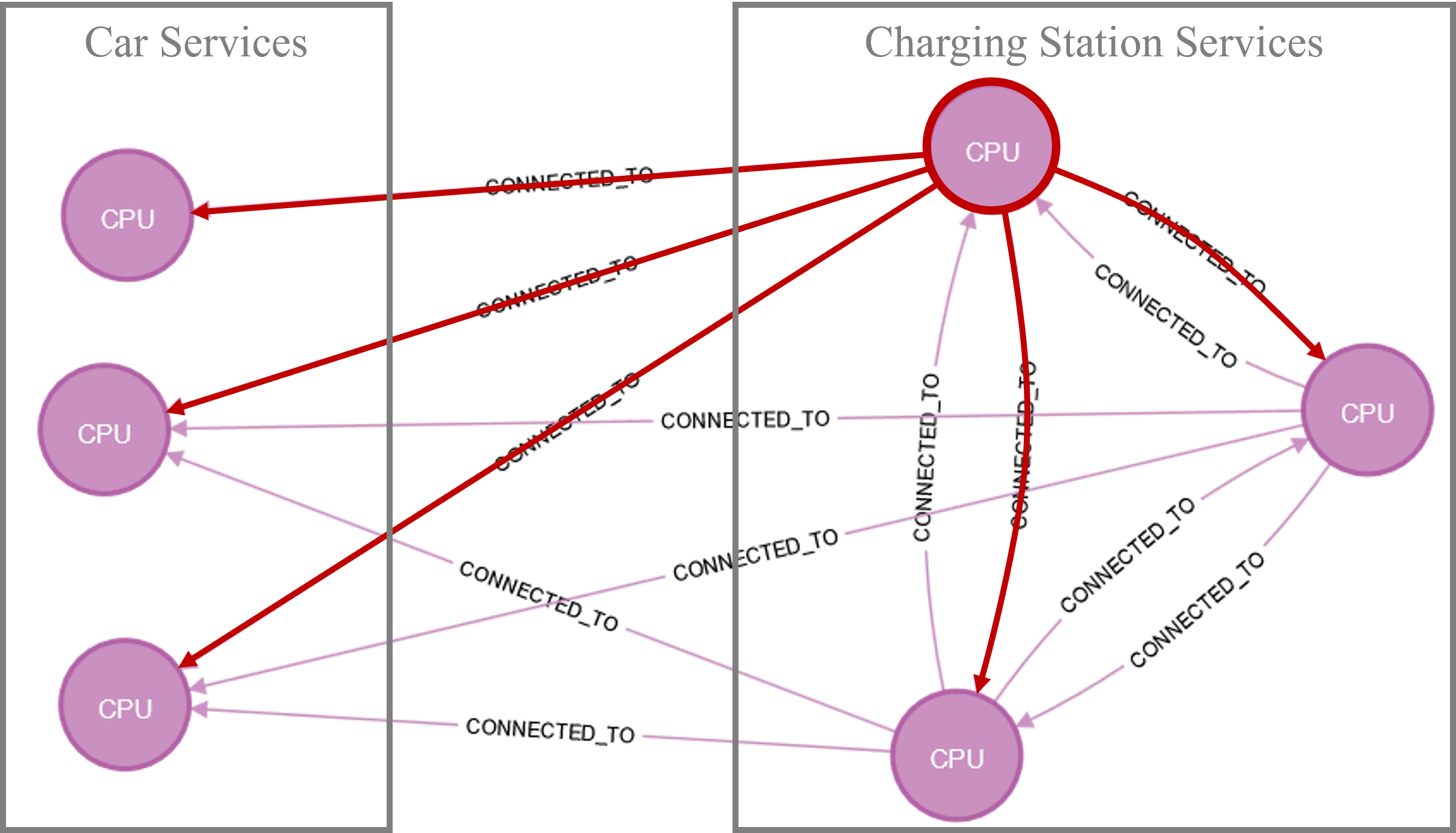}
    \caption{Anomalous causality graph for the charging station test scenario. The increased CPU usage of a single charging station service causes CPU fluctuations across the system.}
    \label{fig:rca_eval}
\end{figure}

Fig.~\ref{fig:rca_eval} presents the resulting anomalous subgraph (service nodes are hidden for clarity). The graph reveals a dense network of causal relationships among the affected metrics, which is expected given the small-scale demo system and the load balancing mechanisms that operate across all worker nodes. Despite this complexity, SDVDiag successfully isolates the root cause of the incident, which can be attributed to the accurate causal discovery model and graph pruning procedure.

Overall, the evaluation demonstrates that SDVDiag is capable of reliably detecting anomalies and identifying their root causes in a dynamic, distributed vehicle environment. The platform's modular architecture, adaptive learning capabilities, and graph-based analysis techniques contribute to a robust and effective diagnosis process.

\section{Conclusion}

To enable the diagnosis of the challenging environment posed by software-defined vehicles, this paper presents SDVDiag as a modular and scalable platform that improves the reliability of connected vehicle systems by combining distributed tracing and causality mining. The platform consists of subsystems for the creation of dependency and causal graphs, a learning environment for continuous model refinement, and incident analysis. Its service-based design enables real-time diagnostics, anomaly detection, and root-cause analysis. Evaluation within a 5G test environment confirms its effectiveness in monitoring applications, providing a robust anomaly detection, and ensuring scalable diagnostics for intelligent mobility. The main findings can be summarized as follows:
\begin{itemize}
    \item Diagnosing failures in distributed vehicle systems is challenging due to complex interdependencies and limited system transparency.
    \item SDVDiag employs distributed tracing and causality mining to map dependencies and identify root causes with precision.
    \item The platform provides effective mechanisms for continuous model adaptation on live operational data, maintaining robust performance despite frequent system changes.
\end{itemize}

While SDVDiag demonstrates satisfactory performance in diagnosing incidents observable through system metrics, many real-world failures require additional expert knowledge for accurate identification. Therefore, future work will focus on integrating advanced causal discovery techniques to reduce training overhead and enable the seamless incorporation of domain expertise into the causality mining process. Additionally, the human-assisted feedback loop for fine-tuning anomaly detection and causal inference will be further developed and comprehensively addressed in future work.


\bibliographystyle{IEEEtran}
\bibliography{bib}

\end{document}